\documentclass{ws-procs975x65}
\newcommand{\be}{\begin{eqnarray}}
\newcommand{\ee}{\end{eqnarray}}

\begin{document}

\title{The concept of correlated density and its application}

\author{Klaus Morawetz$^{1,2}$, Pavel Lipavsk\'y$^{3,4}$, Jan Kol\'a\v cek$^4$, Ernst Helmut Brandt$^5$, Michael Schreiber$^1$}
\address{$^1$Institute of Physics, Chemnitz University of Technology, 
09107 Chemnitz, Germany\\
$^2$Max-Planck-Institute for the Physics of Complex
Systems, Noethnitzer Str. 38, 01187 Dresden, Germany\\
$^3$ Faculty of Mathematics and Physics, Charles University, 
Ke Karlovu 3, 12116 Prague 2, Czech Republic\\
$^4$ Institute of Physics, Academy of Sciences, 
Cukrovarnick\'a 10, 16253 Prague 6, Czech Republic\\
$^5$Max Planck Institute for Metals Research, 70506
Stuttgart, Germany\\
E-mail: morawetz@physik.tu-chemnitz.de, lipavsky@fzu.cz}

\begin{abstract}
The correlated density appears in many physical systems ranging from dense interacting gases up to Fermi liquids which develop a coherent state at low temperatures, the superconductivity. One consequence of the correlated density is the Bernoulli potential in superconductors which compensates forces from dielectric currents. This Bernoulli potential allows to access material parameters. Though within the surface potential these contributions are largely canceled, the bulk measurements with NMR can access this potential. Recent experiments are explained and new ones suggested. The underlying quantum statistical theory in nonequilibrium is the nonlocal kinetic theory developed earlier.
\end{abstract}


\bodymatter

\section{Introduction}\label{}

The theory of dense quantum gases distinguishes the number of free particles and the number of particles in a correlated state, e.g. in bound states. Therefore any kinetic theory should result in a balance equation which in equilibrium leads to the Beth-Uhlenbeck equation of state  \cite{BU36,BU37,SRS90} where the density of free and correlated particles are distinguished $n=n_f+n_{\rm corr}(n_f)$. In contrast to that the Landau theory of quasiparticles describes dense interacting Fermi systems only in terms of the quasiparticle density. This is supported by the Luttinger theorem \cite{L60} which states that the Fermi momentum for any correlated Fermi system in the ground state is exclusively determined by the total number of particles and there is no shift due to correlations. This apparent contradiction between the two approaches is solved within the nonlocal kinetic theory which unifies both pictures and provides them as limiting cases of a more general treatment \cite{LSM97}. How is this possible? First of all an ideal Fermi liquid obeys the Luttinger theorem and any correlated density has to vanish in the ground state. This is in a sense a tautology \cite{F99}. But one has to observe that most Fermi systems do not turn to ideal Fermi liquids at low temperatures. Electrons in metals and nucleons, e.g. develop  a coherent state, the superconducting state. There is a small fraction of correlated density in the superconducting state. Such correlated density shifts the Fermi momentum and consequently the chemical potential $\mu$.  Since the electrochemical potential has to remain constant there must be an electrostatic potential $\phi$ compensating this shift according to $\mu+e\varphi=0$. This electrostatic potential turns out to be of the form of a Bernoulli potential in superconductors. If this potential could be measured we have a proof of the concept of correlated density. 

This overview discusses the theoretical treatments of the Bernoulli potential and the experimental realizations to measure it. First we describe shortly how the correlated density appears in superconducting states and how it is related to the electrostatic potential. Then we review the history of this Bernoulli potential and explain why surface measurements have not been successful in the past. We found the reason in the Budd-Vannimenus theorem which explains why large-scale cancellations appear on the surface of superconductors. Then we calculate what potential can be measured deeper in the bulk by calculating the density and magnetic profile of superconductors. This allows us to explain a recent NMR experiment and to predict inhomogeneous fields to be measured above vortices in type-II superconductors. Finally we conclude with a short description of the nonlocal kinetic theory which provides the description of the correlated density in nonequilibrium.

\section{The concept of correlated density}\label{}

\subsection{Correlated density within BCS approach}

In superconductors the Wigner distribution function  has a two-part structure \cite{A69}
\be
f_W&=&\frac 1 2 \left (1+{\xi \over E}\right ) f(E)+\frac 1 2 \left (1-{\xi \over E}\right ) f(-E)
=\frac 1 2 -{\xi \over 2 E} \tanh {\frac 1 2 \beta E}
\label{f}
\ee
where $\xi=\epsilon_p-\mu$ is the free-particle energy $\epsilon_p$ minus the chemical potential $\mu$. The quasiparticle energy $E=\sqrt{\xi^2+\Delta^2}$ describes the influence of the superconducting gap $\Delta$ on the excitation spectrum of the superconducting state. The Fermi-Dirac distribution is $f(x)=1/({\rm e}^{\beta x}+1)$ with the inverse temperature $\beta=1/k_B T$. There are two branches of energies. For states inside the Fermi sphere, $\xi<0$, the negative branch given by the second term dominates. For the states above the Fermi sphere the role of the contributions is reversed. Close to the Fermi sphere both contributions are comparable.

The density $n$ is obtained by the momentum integral over (\ref{f}) and introducing the density of states $h(\xi)=2 \sum_p 2 \pi \delta (\xi-\epsilon_p)$, one has
\be
n=\int\limits_{-\bar \mu}^\infty {d \xi \over 2 \pi} h(\bar \mu +\xi) \left (\frac 1 2 -{\xi \over 2 E} \tanh {\frac 1 2 \beta E}\right ).
\label{n}
\ee 
Here we account for a possible electrostatic potential $\varphi$ and the velocity $v$ of superconducting electrons by $\bar \mu=\mu -e \varphi-mv^2/2$. For a vanishing gap we obtain the corresponding density $n_n$ of normal electrons with the chemical potential $\bar \mu$ by $n_n=n(\Delta=0)$. The difference 
\be
n_{\rm corr}=n-n_n
\label{nc1}
\ee
describes the correlated density. Let us discuss it first in the ground state. There the normal density turns into
\be
n_n=2 \sum_p \Theta(\bar \mu-\epsilon_p) \approx n_0-(e \varphi+\frac m 2 v^2) {h(\mu) \over 2 \pi}
\label{nn}
\ee
where we have expanded $\bar \mu$ in first order around the Fermi energy and $n_0$ describes the number of particles with no motion and no electrostatic potential. 
The correlated density (\ref{nc1}) splits into two parts in the zero-temperature limit of (\ref{n})
\be
n_{\rm corr}=\frac 1 2 \int\limits_0^\infty {d\xi \over 2 \pi}h(\bar \mu +\xi){\sqrt{\xi^2+\Delta^2}-\xi\over \sqrt{\xi^2+\Delta^2}}-
\frac 1 2 \int\limits_{-\bar \mu}^0 {d\xi \over 2 \pi}h(\bar \mu +\xi){\sqrt{\xi^2+\Delta^2}+\xi\over \sqrt{\xi^2+\Delta^2}}
\ee
which vanishes for vanishing gap. Since the gap is only nonzero in the vicinity of the Fermi level given by the Debye frequency $\omega_D$ we can restrict the integration to the $\pm \omega_D$-range. Expanding the density of states for $\xi<\omega_D$ we obtain finally
\be
n_{\rm corr}={\partial h\over \partial \mu} {\Delta^2\over 4 \pi} \left [\ln \left ( {\omega_D\over \Delta}+\sqrt{{\omega_D^2\over \Delta^2}+1} \right )-{1\over 1+\sqrt{1+{\Delta^2\over \omega_D^2}}} \right ]\approx {\partial h\over \partial \mu} {\Delta^2\over 4 \pi} \ln \left ( {2 \omega_D\over \sqrt{e} \Delta}\right )
\label{nc}
\ee
for $\omega_D\gg \Delta$ in the last step.

Since the total system should stay neutral we expect $n=n_0$ and the two contributions, $n_n-n_0$ according to (\ref{nn}) and $n_{\rm corr}$ of (\ref{nc}), should cancel. Therefore the required electrostatic potential must read 
\be
e\varphi=-\frac m 2 v^2 +{\partial \ln h\over \partial \mu} {\Delta^2 \over 2 } \ln \left ( {2 \omega_D\over \sqrt{e} \Delta} \right ).
\label{fcorr}
\ee
We see that it has the form of a Bernoulli potential which compensates the contribution of the supercurrent velocity and the associated inertial and Lorentz forces and has a part directly linked to a material parameter namely the gap. The latter is called thermodynamic corrections. This internally developed Bernoulli potential and field keep the charge neutrality and compensate the inhomogeneities due to diamagnetic currents. 

The great hope is now to measure the Bernoulli potential in order to access directly the gap parameter. 

\subsection{Bernoulli potential}

A way to discuss the Bernoulli potential closely connected to experimental facts is the London theory. The London condition
\be
m{\bf v}=-e{\bf A}
\ee
fixes the trajectories of superconducting electrons due to the transverse vector potential. This condition explains the Meissner effect and contains both London equations. One can consider it as the simplest way to account for basic electrodynamic features of superconductors. The time derivative of this London condition gives the force 
\be
m\dot{\bf v}=-{ e{\partial{\bf A}\over\partial t}}-{ e({\bf v}
\nabla){\bf A}}
={ e({\bf E}}+{ {\bf v}\times{\bf B}})+\nabla\left({ e
\varphi}+{ {1\over 2}mv^2}\right)
\ee
exerted on the electrons. Comparing this with Newton's equation of motion $m\dot{\bf v}=e({\bf E}+{\bf
  v}\times{\bf B})+{\bf F}_s$
where the unknown force on the superconducting pair is denoted with ${\bf F}_s$
one obtains
\be
\nabla e\varphi={\bf F}_s-\nabla{1\over 2}mv^2.
\label{Lc4}
\ee
The different expressions in literature differ by the different treatment of the unknown force ${\bf F}_s$. London neglected this force consistent with the London condition. This corresponds to the picture of hydrodynamics of a charged ideal gas  and it results in $e\varphi=-\frac 1 2 m v^2$. The diamagnetic current is maintained by the electrostatic potential \cite{B37}. Sorkin \cite{So49} suggested that the free energy responsible for superconductors must contribute, too, which yields $ e \varphi=-\frac 1 2 m v^2 -{\partial f_s\over \partial n_s}$. As a further step van Vijfeiken and Staas \cite{VS64} considered the force resulting from the
interaction between electrons and condensate
acting on
electrons to keep them at rest ${\bf F}_n+e{\bf E}=0$. Taking into account the balance of forces  $n_n{\bf  F}_n+n_s{\bf F}_s=0$ as well as that $e{\bf E}=-e\nabla \varphi$ the force becomes ${\bf F}_s=-{n_n\over
  n_s}e\nabla\varphi$. Introducing this into (\ref{Lc4}) one obtains that the Bernoulli potential is reduced by the fraction of superconducting electrons $e\varphi=-{n_s\over n}\frac 1 2 m v^2$ called quasiparticle screening or fountain effect. Rickayzen \cite{Ri69} finally started from the kinetic energy density $f_{kin}=n_s \frac 1 2 m v^2$ of the condensate which determines
the change in the chemical potential 
\be
e\varphi=-\mu =-{\partial \over \partial n} f_{kin}
=
-{\partial
  n_s\over \partial n} \frac 1 2 m v^2
=-{n_s\over n} \frac 1 2 m v^2
+4 {n_n\over n} {\partial \ln T_c\over \partial \ln n} \frac 1 2 m v^2.
\label{tcorr}
\ee
This expression provides, besides the quasiparticle screening, also thermodynamic corrections. This offers the experimental possibility to access directly material parameters similar to the BCS expression (\ref{fcorr}).

The experimental attempts to measure it, however, have yielded no result. Since Ohmic contact measurements cannot provide a result due to the constant electrochemical potential \cite{Le53,Hu66}, Bok and Klein \cite{BK68} and later more precisely Morris and Brown \cite{MB71} measured the Bernoulli potential at the surface via capacitive pickup. For the investigated Pb at $7$K the thermodynamic corrections have been expected to be larger by a factor 30 than the first term in (\ref{tcorr}). Surprisingly the observed data agree only with the first term of (\ref{tcorr}).  

Why no signal of thermodynamic corrections is seen remained a puzzle for nearly 30 years. Recently we found the solution by a modification 
\cite{LKMM01} of the Budd-Vannimenus theorem \cite{BV73}
\be
\rho_{\rm lat}\left(\varphi_0-\varphi(0)\right)&=&
f_{\rm el}-n{\partial f_{\rm el}\over\partial n}=-n^2 {\partial\over \partial n}\left ( {f_{\rm el} \over n}\right ),
\label{e1}
\ee
which states that the extension of the surface profile towards the surface, $\varphi(0)$ has a step at the surface due to the surface dipoles which is given exclusively in terms of the electronic free energy with no regard to the potential inside. Writing the change of the free energy as Rickayzen, 
$e\varphi=-{\partial \over \partial n} \delta f$ and $\delta f=n_s \frac 1 2 m v^2$, the potential at the surface becomes
\be
e\varphi_{\rm surf}=e \varphi +{ n {\partial \over \partial n} {\delta f\over
  n}}=e\varphi+{\partial \over \partial n} {\delta f}-{\delta f\over n}=-{ { n_s\over n}}\frac 1 2 m v^2.
\ee
The surface dipoles compensate the thermodynamic corrections exactly \cite{LKMM01}.

This is of course strictly valid only for homogeneous superconductors. Even the appearance of vortices in type-II superconductors requires a more detailed discussion. In order to describe such inhomogeneous situation we use the extension of the Landau-Ginzburg equation towards lower temperatures by Bardeen \cite{B54}.

\subsection{Inhomogeneous superconductors within extended Ginzburg-Landau approach}

We consider the free energy as composed of the condensation energy, the kinetic energy, the electrostatic energy as well as the magnetic energy, ${\cal F}[{ \psi,{\bf A},n_n}]=\int d r f={\cal F}_s+{\cal F}_{\rm kin}+{\cal F}_C+{\cal F}_M$. The condensation energy is constructed according to the two-fluid model of Gorter and Casimir \cite{GC34}, where one describes two experimental facts, firstly that the fraction of superconducting electrons shows a temperature dependence of nearly ${ \varpi}=1-{T^4\over T_c^4}\approx {n_s\over n_s+{ n_n}}$ and secondly that the condensation energy is an expression containing the specific heat parameter $\gamma$ and the critical temperature as $\varepsilon_{\rm con}={1\over 4}\gamma T_c^2$. Eliminating the critical temperature one sees that the condensation energy must have the form $\varepsilon_{\rm con}={\gamma T^2\over 4\sqrt{1-{ \varpi}}}$ in equilibrium. This relation should result from a variation of the free energy $\delta{\cal F}_s / \delta\varpi = \partial f_s / \partial \varpi =0$ such that
\be
f_s=U-\varepsilon_{\rm con}{ \varpi}-\frac 1 2\gamma T^2
\sqrt{1 - { \varpi}}.
\ee
The kinetic part is taken as a form proposed by Ginzburg and Landau \cite{GL50}, who suggested to represent the superconducting density by a pseudo wave function $|{ \psi}|^2={n_s\over 2}$ such that the kinetic energy can be given in Schr\"odinger form ($m^*=2 m_e$, $e^*=2 e$)
\be
{\cal F}_{\rm kin}=\int d{\bf r}{1\over 2m^*}
\left|\left(-i\hbar\nabla-e^*{ {\bf A}}\right){ \psi}\right|^2.
\label{FK}
\ee

Variation of the total free energy with respect to $\psi^*$ 
yields the Ginzburg-Landau equation (GL) for the wave function 
\be
{1\over 2m^*}(-i\hbar\nabla-e^*{\bf A})^2\psi+\chi\psi=0
\label{GL}
\ee
with the potential
\be
\chi={\partial f_s\over\partial|\psi|^2}-2
{\partial f_s\over\partial n_n}=-2{\varepsilon_{\rm con}\over n}+{\gamma T^2\over 2n}
{1\over\sqrt{1-{2|\psi|^2\over n}}}.
\label{chi1}
\ee
This extends the GL equation towards lower temperatures. 
Close to $T_c$ the potential takes the form 
$ \chi\to\alpha+\beta|\psi|^2$ and the original GL equation is recovered.

The GL equation was solved numerically\cite{LKMB01} for Nb; the results are seen in figure~\ref{fig1}. The vortices arrange themselves in the hexagonal structure of an Abrikosov-type lattice. The superconducting density is zero at the vortex centers and reaches nearly the nonmagnetic value at the borders of the vortex. The fact that $n_s$ is smaller there than its nonmagnetic value is a result of the nonlocal effects. The corresponding magnetic field reaches its maximum $B_{\rm max}$ at
the vortex centers. In fact, it is higher than the applied field in the core which means that the superconductor compresses the magnetic field in vortices.

\def\figsubcap#1{\par\noindent\centering\footnotesize(#1)}
\begin{figure}[h]%
\begin{center}
  \parbox{2.1in}{\epsfig{figure=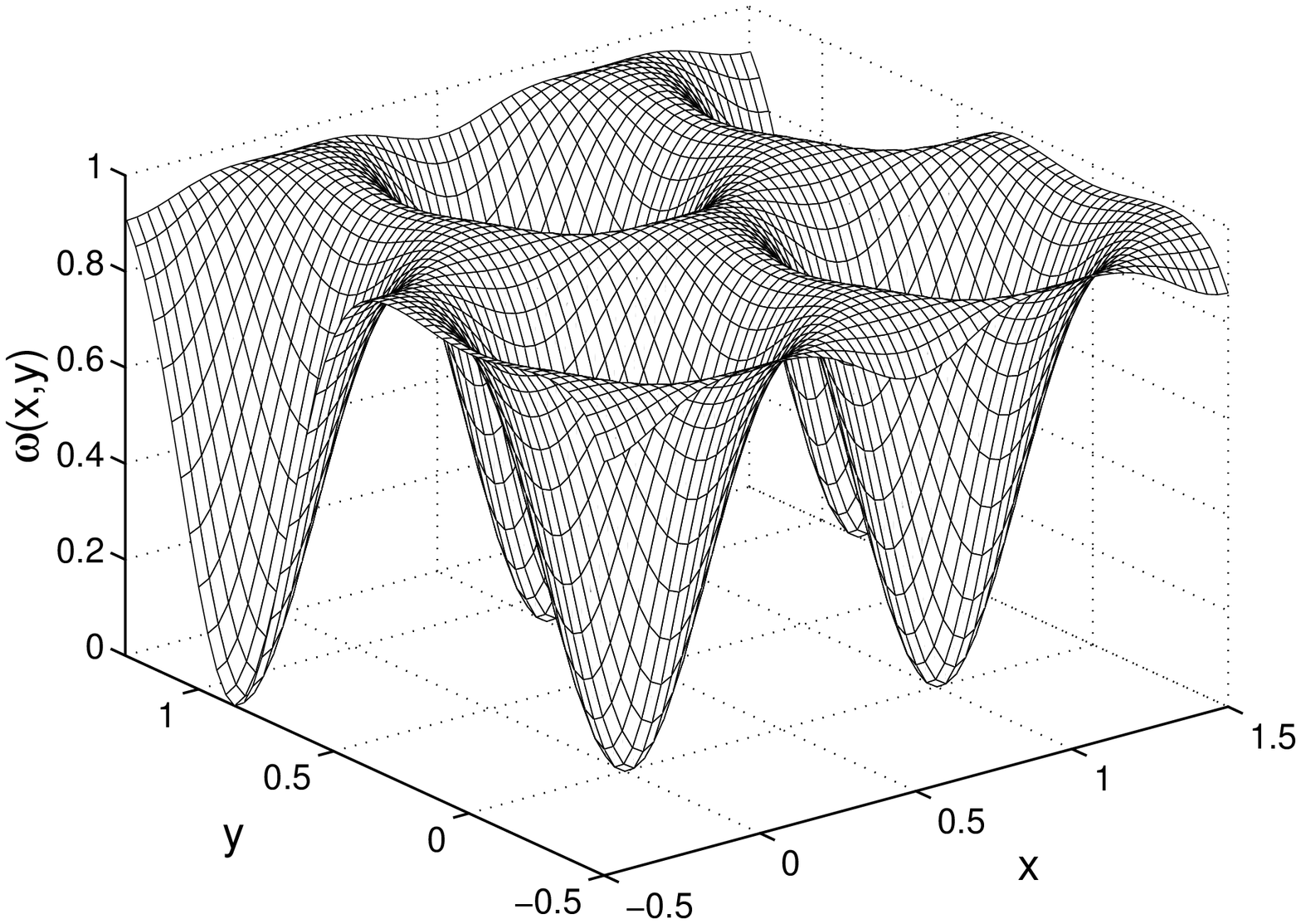,width=2in}\figsubcap{a}}
  \hspace*{4pt}
  \parbox{2.1in}{\epsfig{figure=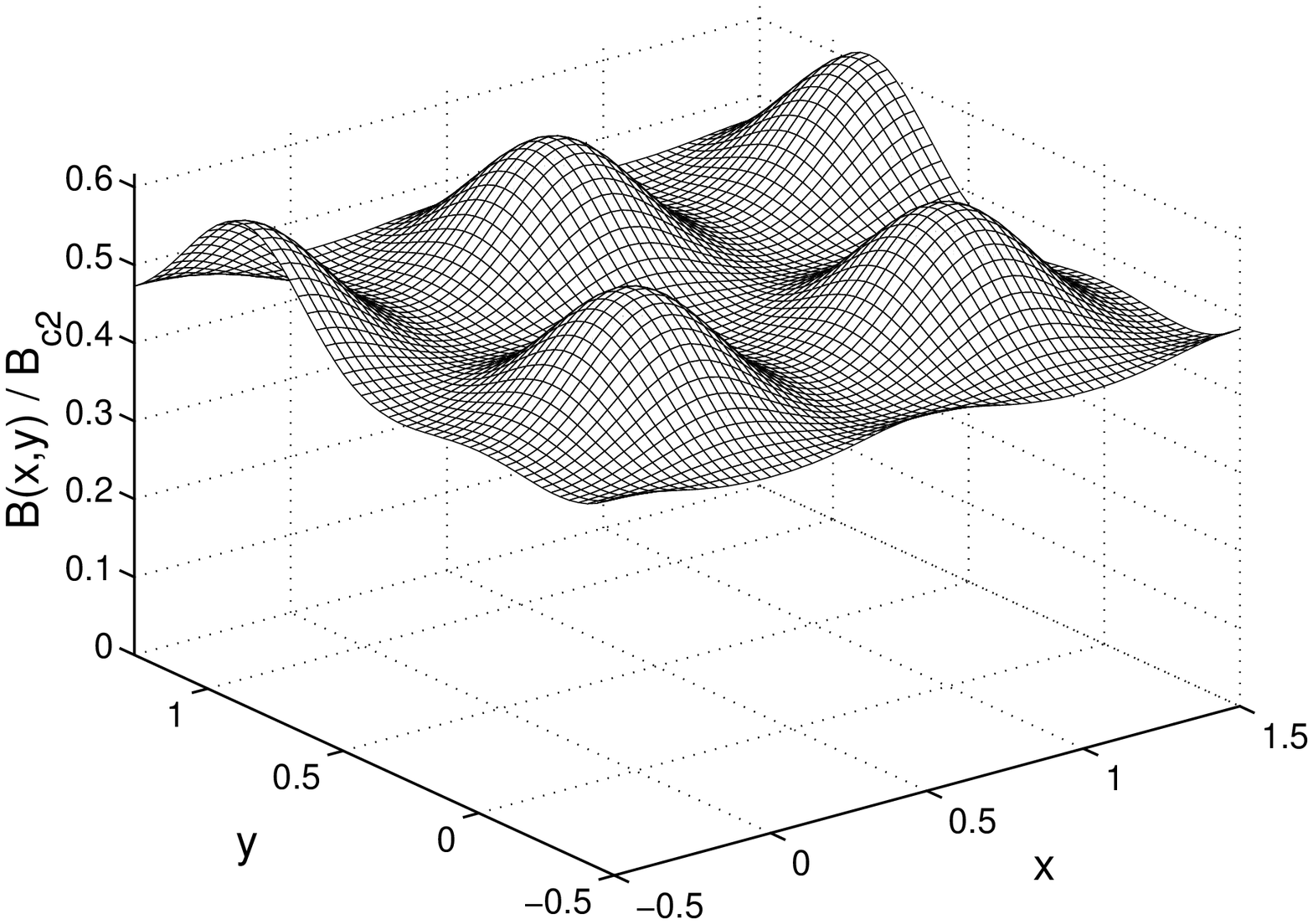,width=2in}\figsubcap{b}}
  \caption{(a) Condensate fraction $|\psi /\psi_0|^2$ normalized to the magnetic-field-free value $\psi_0$ for $T/T_c=0.5$,
the magnetic induction $\bar B/B_{c2} =0.5$, and the GL parameter
$\kappa_0=1.5$. (b) The magnetic field in units of the upper critical field
$B_{c2}$. \protect\cite{LKMB01}}%
  \label{fig1}
\end{center}
\end{figure}

\subsection{Charge profile}

The numerical solution of the GL equation allows us to discuss the detailed charge density and potential profile towards the surface. For magnetic fields perpendicular to the surface we found a linearized solution of the GL equation (\ref{GL}) with respect to low magnetic field $\psi=\sqrt{n(1-(T/T_c)^4)/2}+\delta\psi$
with \cite{LMKMBS04}
\be
\delta\psi={\psi_\infty\over 2-\kappa^2}
{e^{*2}\lambda^4B^2\over 2\hbar^2}\left(
{\rm e}^{-{2x/\lambda}}-{\sqrt{2}\over\kappa}
{\rm e}^{-{\sqrt{2}x/\xi}}\right).
\label{a12}
\ee
where $\lambda=m^* /\mu_0 n e^2$ is the London penetration depth indicating how far the magnetic field penetrates the superconductor, $\xi=\hbar^2/2 m^* \alpha$ is the GL coherence length in terms of $\alpha=\gamma T_c(T-T_c)/n$, and the ratio between both quantities is the GL parameter $\kappa$. One sees that the two length scales determine the profile corresponding to whether $\kappa$ is larger or smaller than $\sqrt{2}$, i.e. type II or type I superconductors. The resulting charge profile results from the variation of the free energy with respect to the normal density and reads
\be
e\delta\varphi&=&-{1\over 2}mv^2
\left({n_s\over n}+4{n_n\over n}{\partial\ln T_c\over\partial
\ln n}\right)
{1\over 1+{\sqrt{2}\over\kappa}}\left(
1+{{\rm e}^{\left({2\over \lambda}-{\sqrt{2}\over \xi}\right)x}-
1\over 1-{\kappa\over\sqrt{2}}}\right).
\label{a21}
\ee
This result remains finite at the limiting case of $\kappa=\sqrt{2}$ and 
yields Rickayzen's result (\ref{tcorr}) for $\kappa \to \infty$.

With the help of the Poisson equation the linearized density profile reads \cite{LMKMBS04}
\be
\rho={2e\epsilon_0 B^2\over m\left(1\!-\!{2\over\kappa^2}\right)}
\left({n_s\over n}\!+\!4{n_n\over n}
{\partial\ln T_c\over\partial\ln n}\right)\!\!
\left({\rm e}^{\!-\!{2x/\lambda}}\!-\!{\kappa\over\sqrt{2}}
{\rm e}^{\!-\!{\sqrt{2}x/\xi}}\right)\!.
\label{a23}
\ee
The surface value, $\rho(x=0)$, is always negative while the 
bulk value, $\rho(x\gg 0)$, remains 
positive. This allows us to define the 
width of the surface by $\rho(w)=0$ with the result
\be
w={\lambda\over 2}{\ln{\kappa\over\sqrt{2}}\over
{\kappa\over\sqrt{2}}-1}.
\label{a25}
\ee
For type-II superconductors, $\kappa>\sqrt{2}$, the surface charge is formed by the contribution on the scale of the 
GL coherence length $\xi$
while for 
$\kappa \to \infty$ the width $w\to {\xi\over\sqrt{2}}\ln{\kappa\over\sqrt{2}}$ and the bulk charge extends on the scale of $\lambda$.
For type-I superconductors, $\kappa<\sqrt{2}$, the situation is reversed, the  surface charge is formed by the contribution on the scale of the London penetration
depth $\lambda$
and for
$\kappa \ll \sqrt{2}$ the width $w\to {\lambda
\over 2}\ln{\sqrt{2}\over\kappa}$  and the bulk charge extends on the scale of $\xi$.

\section{Application of the Bernoulli potential}
\subsection{Surface potential}
Summarizing, the electrostatic potential can leak out of a superconductor by three types of charges:
(i) The bulk charge which describes the transfer of electrons from the inner to the outer regions of vortices creating a Coulomb 
force. This force has to balance the centrifugal force by the electrons rotating around the vortex center, the outward push of the magnetic field 
via the Lorentz force and the outward force coming from the fact that the energy of Cooper pairs is lower than the one of free 
electrons such that unpaired electrons in the vortex core
are attracted towards the condensate around the core \cite{LMKMBS04}.
(ii) The surface dipole which cancels all contributions of pairing 
forces \cite{LMKMBSa04} resulting in an 
observable surface potential of
\be
e\phi_0=-{f_{\rm el}\over n}.
\label{BV}
\ee 
(iii)
The surface charge \cite{LMKMBSb04}
distributed on the scale of the Thomas-Fermi screening length $\lambda_{\rm TF}$ with $\xi, \xi_0, \lambda \gg\lambda_{\rm TF} \rightarrow 0$. 

Within the discussed Bardeen's extension of the GL equation the free electron energy has three components

\be
f_{\rm el}&=&{ {1\over 2}\gamma T^2}+
{1\over 2m^*}\psi^*\left(-i\hbar\nabla-e^*{\bf A}
\right)^2\psi
{ -\varepsilon_{\rm con}{2|\psi|^2\over n}-
{1\over 2}\gamma T^2\sqrt{1-{2|\psi|^2\over n}}}
\label{ff}
\ee
which determines, according to the modified Budd-Vannimenus theorem (\ref{BV}), the surface potential. Near the critical temperature this becomes very simple $
e\phi_0={1\over 2n}\beta|\psi|^4
$ in terms of $\beta=\gamma T_c^2/2 n^2$.
Without surface dipoles, the surface potential
equals the internal potential
\be
e\phi &=&{ -{1\over 2m^*n}
\psi^*\left(-i\hbar\nabla-e^*{\bf A}\right)^2\psi}
+{ {\partial\varepsilon_{\rm con}\over\partial n}
{2|\psi|^2\over n}}-{T^2\over 2}{\partial\gamma\over\partial n}
\left({|\psi|^2\over n}+{|\psi|^4\over 2n^2}\right)
\ee
where the first part are the {inertial and Lorentz forces} neglecting pairing forces. The approximation by 
Khomskii and 
Kusmartsev \cite{KK92} adopted by Blatter \cite{B96}, 
$
{ e\phi_{\rm Bl}={\gamma T_c\over n}{\partial T_c\over\partial n}
|\psi|^2}
$, takes only the pairing force into account.
Before comparing the different approximations of the literature with our result, it is illustrative to discuss the magnetic field and temperature dependence. For thin layers and a $B$ field close to  $B_{c2}$  the spatially averaged values of the superconducting fraction  $\omega={|\psi|^2 \over |\psi_\infty|^2}$ can be given by general arguments \cite{dG89} as $\langle\omega\rangle\!=\!{1\!-\!b\over \beta_{\rm A}}$ and $\langle\omega^2\rangle\!=\!{(1\!-\!b)^2 \over \beta_{\rm A}}$ with a number $\beta_A$ characteristic for the vortex geometry and $b={B \over B_{c2}}$. Knowing our averaged surface potential we compare it now with Blatter's approximation
\be
\langle e\phi_0\rangle &=&{\varepsilon_{\rm con}\over n \beta_{\rm A}}
\left(1-t^2\right)^2
\left(1-b\right)^2 \\ 
\langle e\phi_{\rm Bl}\rangle&=&{\varepsilon_{\rm con}\over n \beta_{\rm A}}
{\partial\ln T_c\over\partial\ln n}\,2\,(1-t^4)
\left(1-b\right)
\ee
where $t=T/T_c$. Obviously
the surface dipole 
strongly modifies the magnitude of the potential, in particular
when the GL wave function has a small magnitude we have $\phi_0\propto |\psi|^4$, 
while without dipoles $\phi_{\rm Bl}\propto |\psi|^2$.

Possible experimental consequences for which the presented
theory can be tested are
at temperatures close the critical temperature, $t\to 1$, where
$|\psi|^2\propto 1-t$, therefore $\phi_0\propto (1-t)^2$ while
$\phi_{\rm Bl}\propto 1-t$
and at
magnetic fields close to
the upper critical field, $b\to 1$, where $|\psi|^2\propto 1-b$ so that
$\phi_0\propto (1-b)^2$ while $\phi_{\rm Bl}\propto 1-b$.

\def\figsubcap#1{\par\noindent\centering\footnotesize(#1)}
\begin{figure}[h]%
\begin{center}
  \parbox{2.1in}{\epsfig{figure=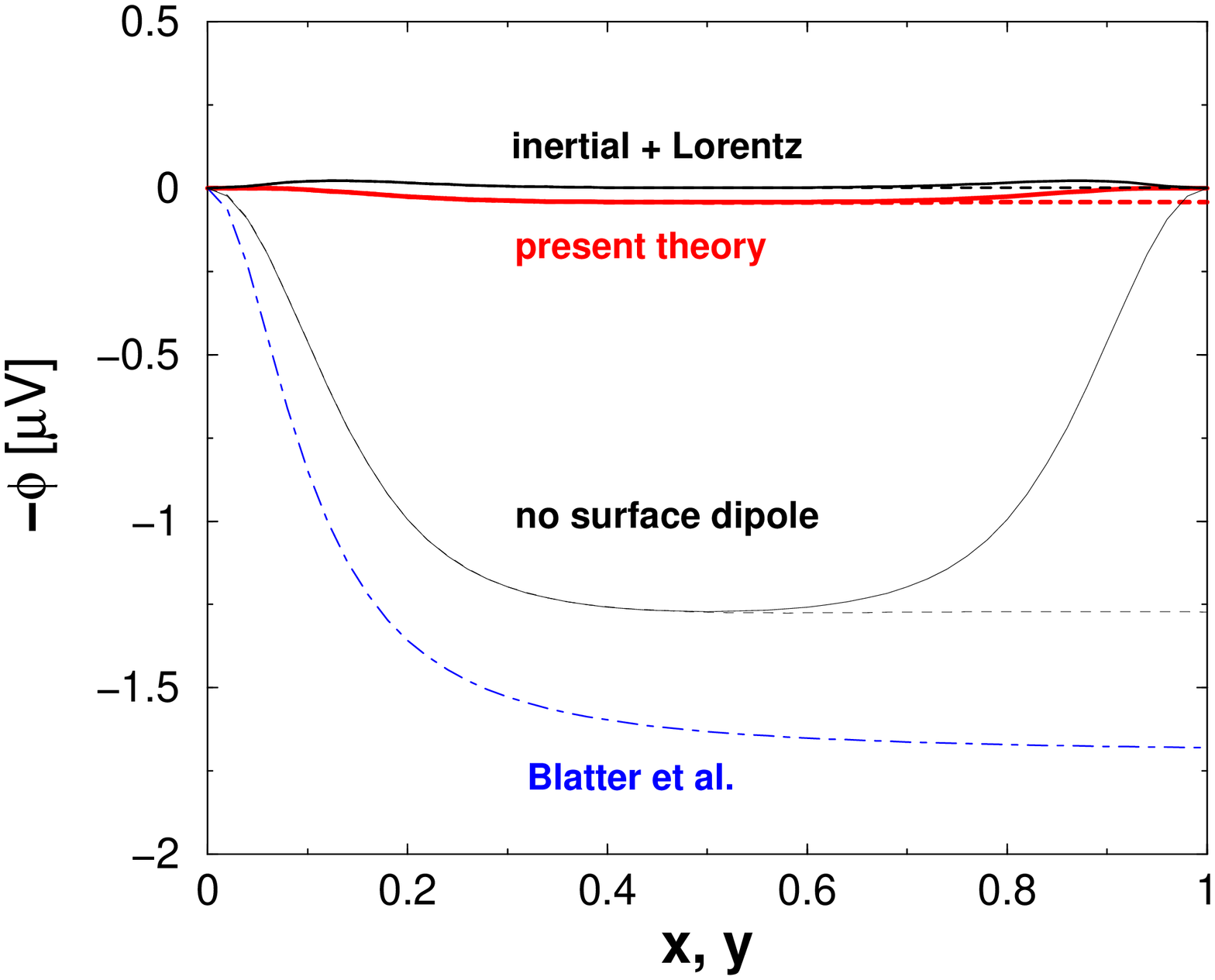,height=2in,angle=-90}\figsubcap{a}}
  \hspace*{20pt}
  \parbox{2.1in}{\epsfig{figure=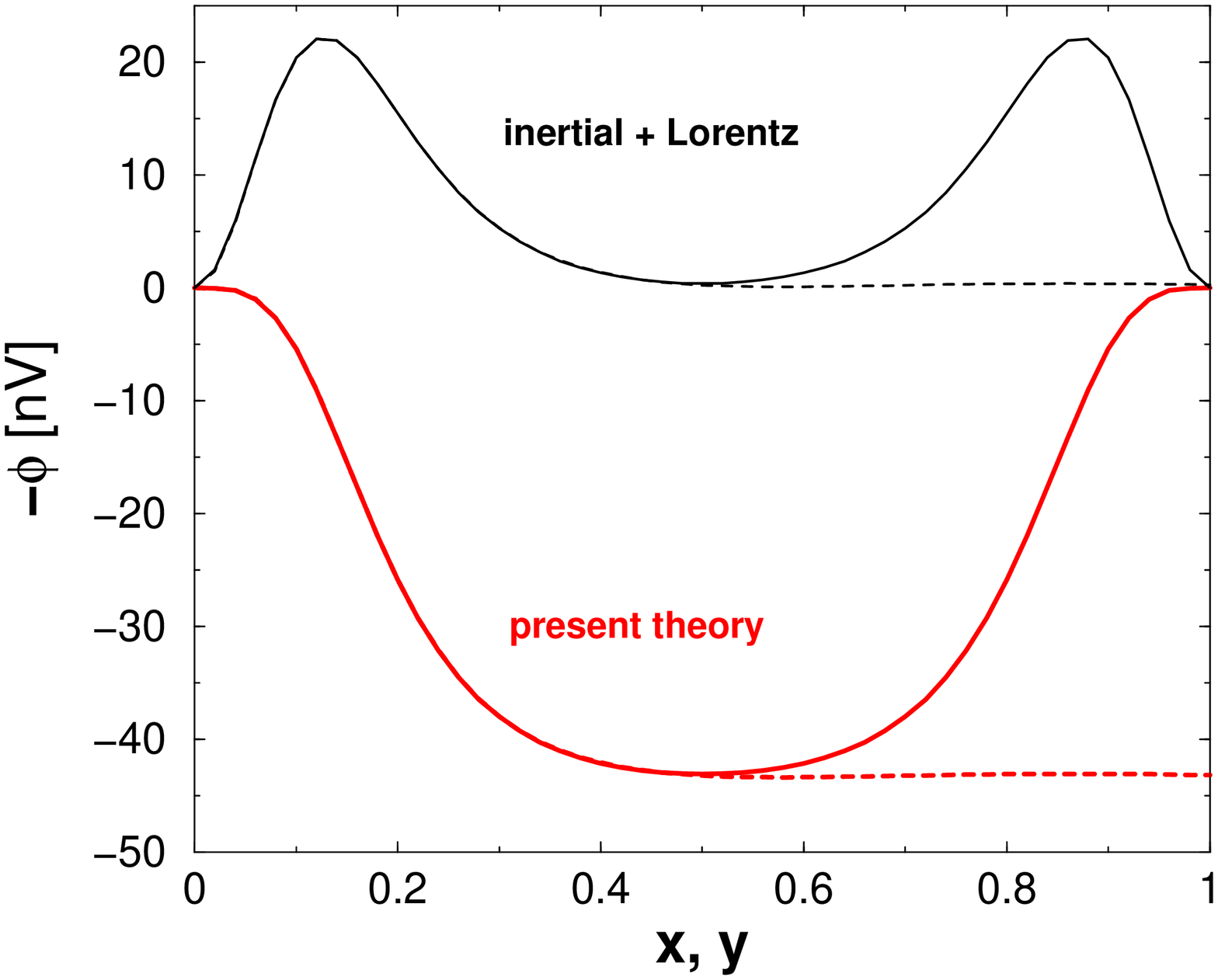,height=2in,angle=-90}\figsubcap{b}}
  \caption{(a) Surface potential for different approximations at ${T\over T_c}=0.95$, $\kappa=0.78$ and ${B\over B_{c2}}=0.7818.$ The broken lines indicate the behavior along $y$ directions different from the $x$ directions due to the hexagonal structure. (b) An enlarged view of the approximation of only inertial and Lorentz forces compared to the present theory.\cite{LMKMBSb04}}%
  \label{fig2}
\end{center}
\end{figure}

The numerical comparison \cite{LMKMBSb04} for Nb is shown in figure~\ref{fig2}. The internal potential and Blatter's result are similar.
The {full theory} and the approximations of only inertial/Lorentz 
forces are much smaller. 
The surface dipole cancels a major part of 
the pairing forces.
The full 
theory and the inertial/Lorentz forces 
even result in different profiles and sign. This illustrates the delicate balance between the three contributions in (\ref{ff}). It is not justified to neglect just one of them. Corresponding results in the literature should be critically reexamined, e.g. the vortex induced deformation and the experimentally accessible effective mass of vortices show remarkably different results \cite{LMKB06}.

\subsection{Charged vortices probed by NMR}

Though the capacitive coupling measurements of the surface potential do not yield an access to thermodynamic corrections and the high precision measurements of the electrostatic potential leaking out of the surface remain to be performed, there is a direct attempt to measure the charge deeper in the bulk by NMR. Kumagai et al. \cite{KNM01} have measured the quadrupole resonance lines in the high-$T_c$ material YBCO. The polarization of the Cu atoms leads to a coupling of spins with the electrical field gradient
and to a splitting of the quadrupole resonance
\be
\nu_Q^{\rm NQR}=E_{\pm 3/2}-E_{\pm 1/2}=A {\rho} +C
\ee 
proportional to the charge density $\rho$. This allows to measure the charge of the vortices. If one advocates the picture that the positive charges are expelled from the vortex to the outer regions one would expect a BCS estimate of the measured charge per length $
Q\approx {2 e k_F  \over \pi^3} \left ( 
{\lambda_{\rm TF}\over \xi}
\right )^2 
\left (
{d \ln T_c\over d \ln \mu}
\right )
$.
This means for the underdoped regime $Q>0$ and for the overdoped $Q<0$, which is just the opposite to what is observed experimentally. Moreover the BCS estimate gives a charge of $10^{-5} e$ per vortex while the experiment yielded a value three orders of magnitude larger. We solved this apparent contradiction with the help of the Bernoulli potential as follows. 

\def\figsubcap#1{\par\noindent\centering\footnotesize(#1)}
\begin{figure}[h]%
\begin{center}
  \parbox{2.1in}{\epsfig{figure=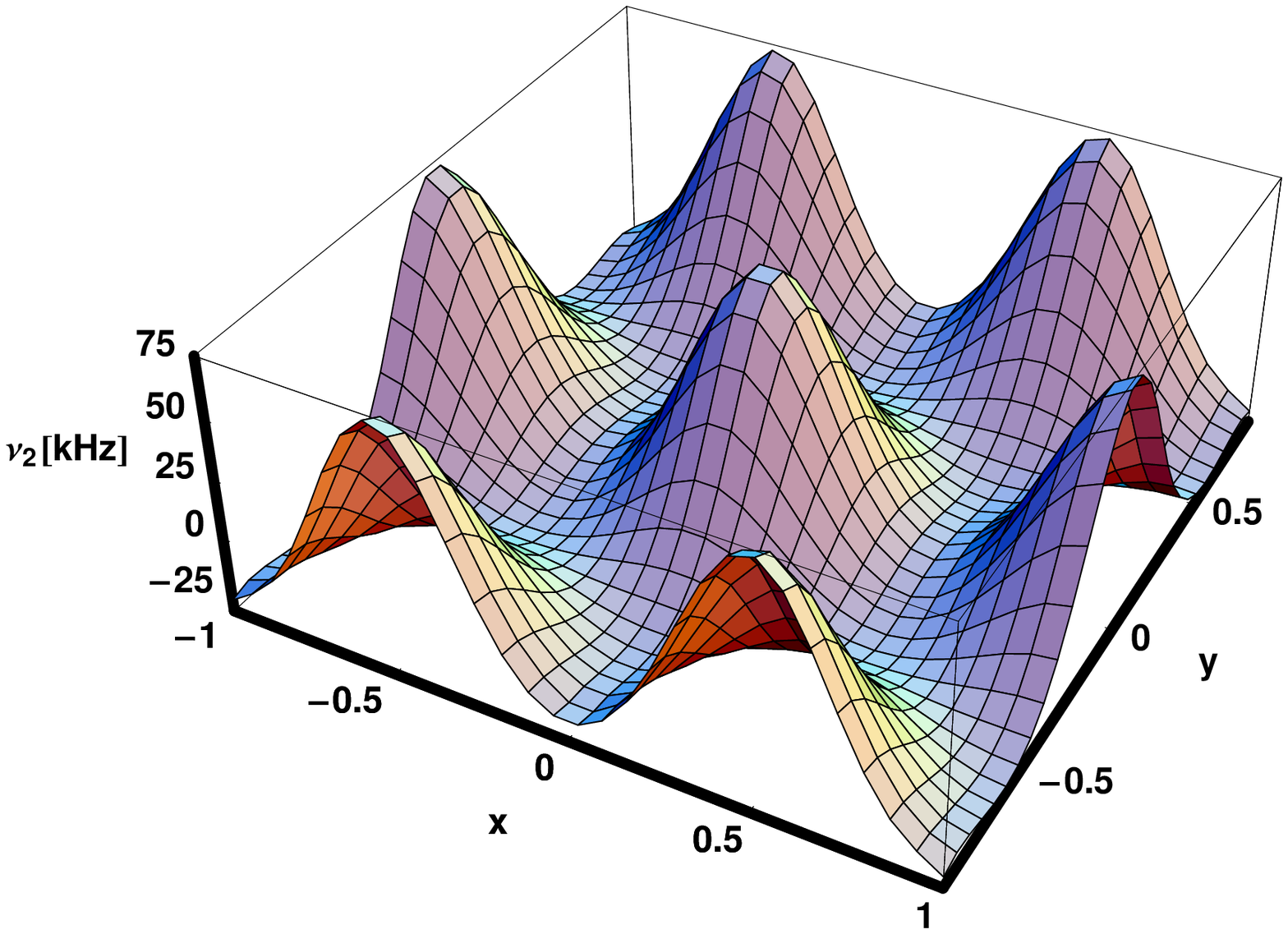,width=2in,angle=0}\figsubcap{a}}
  \hspace*{4pt}
  \parbox{2.1in}{\epsfig{figure=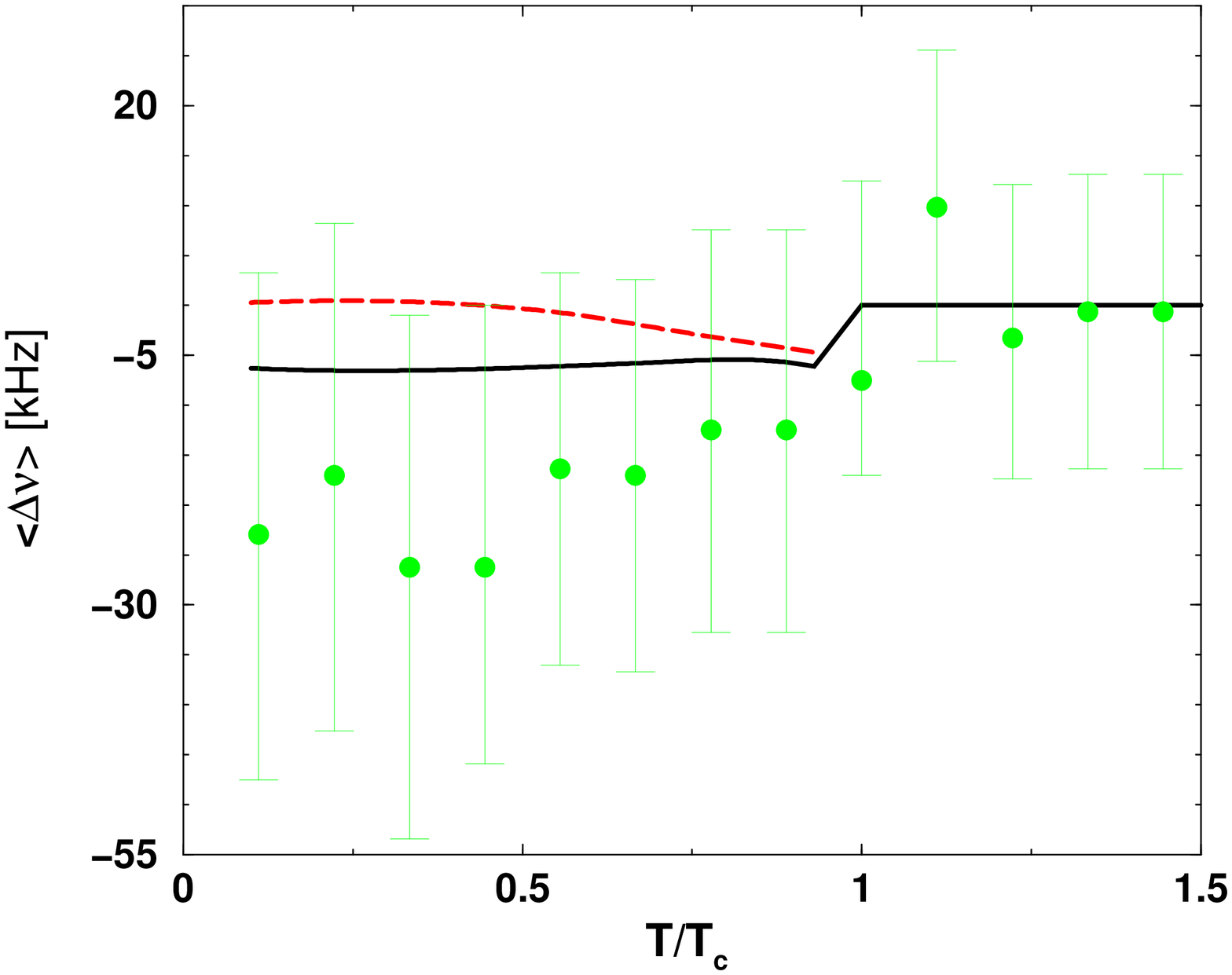,height=2in,angle=-90}\figsubcap{b}}
  \caption{(a) Spatial variation of the line shift $\nu_2=E_{-1/2}-E_{-3/2}$. The units are in vortex sizes. (b) Shift
$\langle\Delta\nu\rangle$ of $\Gamma=140$kHz of a {single crystal} (dashed line), 
and after {averaging over grain orientation} (solid line) compared to the experimental values. \cite{LKMB01}}%
  \label{fig3}
\end{center}
\end{figure}

One has to consider that YBCO is built up from Cu chains and planes where in the latter the superconductivity and the vortices are located. The discussed Bernoulli potential therefore appears in the planes. On the other hand the charge neutrality requires that the exceeding charge has to be compensated in the Cu planes such that a charge transfer appears between planes and chains. Taking this and the two-dimensional geometry into account we have been able to explain the observed quadrupole line shift \cite{LKMB02}, see figure~\ref{fig3}. The experimental line width is comparable to the spatial variation of the charge and magnetic field profile. Therefore the actual spatial profile of the magnetic field and charge density is important as presented in figure~\ref{fig3}.

\section{Consequences to the kinetic theory}

\subsection{Extended quasiparticle approximation}

We have seen that the concept of correlated density leads in a natural way to the Bernoulli potential, which is useful in explaining recent experiments and promising to access a material parameter in superconductors. The underlying nonequilibrium theory has to distinguish between parts of the Wigner distribution described by quasiparticle poles and the rest called off-shell parts.
This 
off-shell motion can be eliminated from the kinetic equation, which
requires to introduce an effective distribution (the quasiparticle
distribution $f$) from which the Wigner distribution $f_W$ can be
constructed
\begin{equation}
f_W[f]=f+\int{d\omega\over 2\pi} {\wp\over\omega-\varepsilon}
{\partial\over\partial\omega}
\left((1-f)\sigma^<_\omega-f\sigma^>_\omega\right).
\label{tr2a}
\end{equation}
Here $\sigma^{>}$ and $\sigma^{<}$ denote the selfenergies describing all correlations and $\varepsilon$ is the quasiparticle energy. This relation represents the extended quasiparticle picture derived for small
scattering rates \cite{SLM96,MLS00,LSM97,Mo04}. 
The limit of small scattering rates has been first introduced by Craig
\cite{C66a}. An inverse relation $f[f_W]$ has been constructed
\cite{BD68}. For equilibrium non-ideal plasmas this approximation has been 
employed by \cite{SZ79,KKL84} and has been used under the name of the generalized
Beth-Uhlenbeck approach by \cite{SR87} in nuclear matter 
for studies of the correlated density. The authors in \cite{KM93} have 
used this approximation with the name "extended quasiparticle 
approximation" for the study of the mean removal energy and high-momenta 
tails of Wigner's distribution. The non-equilibrium form has been derived 
finally as the modified Kadanoff and Baym ansatz \cite{SL95}. We
will call this the extended quasiparticle approximation. 

This extended quasiparticle picture leads to balance equations which include 
explicit correlation parts analogous to the virial corrections. The firmly 
established concept of the equilibrium virial expansion has been
extended to nonequilibrium systems \cite{LSM97} 
although a number of attempts have been made to modify the 
Boltzmann equation so that its equilibrium limit would cover at 
least the second 
virial coefficient \cite{B69,Ba84,S91}. The corrections to 
the Boltzmann equation have the form of gradients or nonlocal contributions 
to the scattering integral.
The nonlocal quasiparticle kinetic
equation for the momentum-, space-, and time-dependent distribution function of particle $a$, $f_1\equiv f_a(k,r,t)$, derived within the non-equilibrium Green's
function technique \cite{SLM96,LSM97} has the form of a Boltzmann equation with the quasiparticle energy $\varepsilon_1=\varepsilon(k,r,t)$,
\begin{eqnarray}
\!\!\!\!\!\!\!\!\!\!\!\!\!&&{\partial f_1\over\partial t}+{\partial\varepsilon_1\over\partial k}
{\partial f_1\over\partial r}-{\partial\varepsilon_1\over\partial r}
{\partial f_1\over\partial k}
=s\sum\limits_b\int{dpdq\over(2\pi)^5\hbar^7}{\cal P}_\pm
\nonumber\\
\!\!\!\!\!\!\!\!\!\!\!\!\!&&\times\Bigl[\bigl(1\!-\!f_1\bigr)\bigl(1\!-\!f_2^-\bigr)f_3^-f_4^--
f_1f_2^\pm\bigl(1\!-\!f_3^\pm\bigr)\bigl(1\!-\!f_4^\pm\bigr)\Bigr]
\label{kin}
\end{eqnarray}
and the spin-isospin etc. degeneracy $s$.
The superscripts $\pm$ denote the signs of non-local corrections:
$f_2^\pm\equiv f_b(p,r\!\pm\!\Delta_2,t)$,
$f_3^\pm\equiv f_a(k\!-\!q\!\pm\!\Delta_K,r\!\pm\!\Delta_3,t\!\pm\!
\Delta_t)$, and $f_4^\pm\equiv f_b(p\!+\!q\!\pm\!\Delta_K,r\!\pm\!
\Delta_4,t\!\pm\!\Delta_t)$. For the out-scattering part of
(\ref{kin}) either the plus or minus signs can be chosen \cite{SLM98}.  
The scattering measure is given by the
modulus of the scattering T-matrix ${\cal P}_\pm=|{\cal T}^R_\pm|^2 \delta (\varepsilon_1+\varepsilon_2-\varepsilon_3-\varepsilon_4\pm 2 \Delta_E)$.
All corrections
$\Delta$, describing the nonlocal and non-instant collision are given by derivatives of the scattering phase shift
\mbox{$\phi={\rm Im\ ln}{\cal T}^R(\Omega,k,p,q,t,r)$}
\begin{equation}
\begin{array}{lclrcl}\Delta_t&=&{\displaystyle
\left.{\partial\phi\over\partial\Omega}
\right|_{\varepsilon_1+\varepsilon_2}},
&\ \ \Delta_2&=&
{\displaystyle\left({\partial\phi\over\partial p}-
{\partial\phi\over\partial q}-{\partial\phi\over\partial k}
\right)_{\varepsilon_1+\varepsilon_2}},
\\ &&&&&\\ 
\Delta_E&=&
{\displaystyle\left.-{1\over 2}{\partial\phi\over\partial t}
\right|_{\varepsilon_1+\varepsilon_2}},
&\Delta_3&=&
{\displaystyle\left.-{\partial\phi\over\partial k}
\right|_{\varepsilon_1+\varepsilon_2}},
\\ &&&&&\\ 
\Delta_K&=&
{\displaystyle\left.{1\over 2}{\partial\phi\over\partial r}
\right|_{\varepsilon_1+\varepsilon_2}},
&\Delta_4&=&
{\displaystyle-\left({\partial\phi\over\partial k}+
{\partial\phi\over\partial q}\right)_{\varepsilon_1+\varepsilon_2}}.
\end{array}
\label{SHIFTSALL}
\end{equation}
The nonlocal kinetic equation (\ref{kin}) covers all quantum virial corrections on the
binary level and conserves density, momentum and energy including the
corresponding two-particle correlated parts \cite{LSM97}. It requires
no more computational power than solving the Boltzmann equation \cite{MLSCN98,MLNCCT01}.

\subsection{Conservation Laws}\label{3.3}

Neglecting all shifts, the time-invariant observables are the 
mean quasiparticle density $n^{\rm qp}$, the mean momentum ${\cal Q}^{\rm qp}$, the mass current $j^{\rm qp}$, the mean energy ${\cal E}^{\rm qp}$ and the stress tensor ${\cal J}_{ij}^{\rm qp}$ in the form
\be
\begin{array}{llllll}
n^{\rm qp}&=&\sum\limits_{k}f,
&\quad
{\cal Q}^{\rm qp}&=&\sum\limits_k k\,f,
~~~~~~~~~~~j^{\rm qp}=\sum\limits_k{\partial
  \varepsilon \over \partial k} f,
\\
{\cal E}^{\rm qp}&=&\sum\limits_{k}({k^2 \over 2 m}+
\frac 1 2 {\sigma_{\rm mf}}) f_k,
&\quad
{\cal J}_{ij}^{\rm qp}&=&\sum\limits_{k}\left(k_j
{\partial\varepsilon\over\partial k_i}+\delta_{ij}\varepsilon\right)f-
\delta_{ij}{\cal E}^{\rm qp},
\label{qlan}
\end{array}
\ee
where we abbreviated the mean-field part $\sigma_{\rm mf}=
\int{dp\over(2\pi)^3}\,t^R(\omega+\varepsilon_p,k,p,0)f_p$ of the 
selfenergy.

Taking now into account the binary correlations which we have 
reformulated into the shifts, we obtain from the non-local kinetic 
equation the modified balance equations \cite{LSM97}
\be
{\partial (n^{\rm free}+{ n^{\rm corr}})\over\partial t}+{\partial
  (j^{\rm qp}+{ j^{\rm corr}})\over\partial r}&=&0
\label{ncorr}\\
{\partial({\cal Q}_j^{\rm qp}+{ {\cal Q}_j^{\rm corr}})\over\partial
  t}+\sum_i{\partial({\cal J}_{ij}^{\rm qp}+{ {\cal J}_{ij}^{\rm corr}})\over
\partial r_i}&=&0\\
{\partial{\cal E}\over\partial t}={\partial 
({\cal E}^{\rm qp}+{ {\cal E}^{\rm corr}}) \over\partial t}&=&0.
\label{ob1}
\ee
This shows that the conserving observables now consist of the sum 
of the quasiparticle parts (\ref{qlan}) and the correlated parts 
which we can interpret as the parts coming from correlated pairs 
or molecules,
\be
&&\begin{array}{llllll}
{ n^{\rm corr}}&=&\int d{\cal P}\Delta_t
&\quad
{ j^{\rm corr}}&=&\int d{\cal P}\Delta_3
\nonumber\\
{ {\cal Q}^{\rm corr}}&=&\int d{\cal P}
{k+p\over 2}\Delta_t &\quad
{ {\cal E}^{\rm corr}}&=&\int d{\cal P}{\epsilon_k+\epsilon_p \over 2}
\Delta_t 
\end{array}
\label{nqcorr}\\
&&{ {\cal J}_{ij}^{\rm corr}}={1\over 2}\int d
{\cal P} \biggl\{k_j\Delta_{3i}+p_j(\Delta_{4i}-\Delta_{2i})+
q_j(\Delta_{4i}-\Delta_{3i})\biggr\},
\label{twoq}
\ee
where $d {\cal P}$ is the probability to form a pair per unit of time. 
This immediate interpretation of $d{\cal P}$ is obvious from the 
expressions (\ref{twoq}). The density of pairs is given, if $d {\cal P}$
is multiplied by the lifetime of the molecule $\Delta_t$. The 
energy, the momentum and mass current which are carried by the pairs 
result analogously. As one can see, the correlated part of the stress 
tensor (\ref{twoq}) takes the form of a virial which is well known for the collision flux in dense
gases. The corresponding momenta are connected to the corresponding 
offsets. The diagonal part of this stress tensor is of course the 
correlated part of the pressure.
The proofs that these conservation laws indeed follow from the nonlocal kinetic equation and are consistent with the extended quasiparticle picture can be found in 
detail in the appendix G of \cite{LSM97}.

\section{Conclusion}
The nonlocal kinetic equation unifying the achievements of transport theory in
dense 
gases with the
quantum transport of dense Fermi systems is presented. The 
quasiparticle drift of Landau's
equation for the evolution of the quasiparticle distribution is connected with a dissipation governed by a nonlocal and 
non-instantaneous
scattering integral in the spirit of Enskog corrections \cite{MLSK99}. 
In this way quantum
transport can be recast into a quasiclassical picture.  
The balance equations for the density, momentum and
energy include quasiparticle contributions and the correlated two-particle contributions beyond the Landau theory \cite{LSM99}.

Compared with the Boltzmann equation, the presented form of virial
corrections only slightly increases the numerical demands in 
implementations. Within INDRA collaboration, we have performed numerical studies and 
compared them with the experimental data from GANIL. The 
nonlocal corrections with parameters found from collisions of two 
isolated nucleons \cite{MLSK99} have been implemented into numerical 
simulations of heavy ion reactions in the non-relativistic regime 
\cite{MLSCN98,MLNCCT01}. Keeping all simulation parameters as in the 
local approximation, the temperature of mono-nucleon products of 
central reactions increases due to nonlocal effects towards 
experimentally observed values, while the temperature of more complex 
clusters remains unchanged \cite{MLSCN98}. For the proton 
distribution in very peripheral reactions we have achieved an 
agreement between theoretical predictions and experimental data 
\cite{MLNCCT01}.

As a consequence of the nonlocal kinetic theory, the correlated density appears which results in the Bernoulli potential in superconductors. This Bernoulli potential allows to access material parameters. Since at the surface large-scale cancellations appear, the charge of vortices as measured by NMR could be explained. We believe the correlated density to be a fruitful concept to describe strong correlations in interacting systems.

\bibliographystyle{ws-procs975x65}
\bibliography{kmsr,kmsr1,kmsr2,kmsr3,kmsr4,kmsr5,kmsr6,kmsr7,delay2,delay3,spin,gdr,refer,sem1,sem2,sem3,micha,genn,short}

\end{document}